\documentstyle[aps]{revtex}
\begin{document}
\title{Flow of low energy couplings in the Wilson
renormalization group}
\author{Robert C. Myers}
\address{Department of Physics, McGill University,
Montr\'eal, PQ H3A 2T8, Canada}
\author{Vipul Periwal}
\address{Department of Physics,
Princeton University,
Princeton, New Jersey 08544}
\maketitle
\begin{abstract} A new form of the Wilson renormalization group
equation is derived, in which the flow equations are, up to
linear terms, proportional to a gradient flow.
A set of co\"ordinates is
found in which the flow of marginal, low-energy, couplings takes
a gradient form, if relevant couplings are tuned to vanish.
\end{abstract}

\def\reef#1{(\ref{#1})}
This paper contains two simple general results on the structure of the
Wilson renormalization group\cite{wil,pol}.   We derive a new form of
the Wilson renormalization group equation, which is similar to a
gradient flow\cite{wally}, and we use this form of the equation to 
show that
there is a set of co\"ordinates such that gradient flow is obtained
for marginal low energy couplings, when relevant couplings vanish.
We emphasize that the flows we consider are {\it not}
beta functions. The difference lies in the treatment of `kinetic'
terms, as we will make precise in our discussion --- see
eqs.~\reef{rescale} and \reef{vanish}.

There has been much work in recent
years on the Wilson approach to the renormalization group\cite{lots},
some of it motivated by Zamolodchikov's beautiful result in two
dimensions\cite{zam,zam2}, and by Cardy's proposal\cite{car} for 
higher
dimensions, and some of it by the string
theoretic insights of Banks and
Martinec\cite{bm}.  More recently, the work on supersymmetric 
Yang-Mills
theories initiated by Seiberg\cite{sei} also makes use of Wilsonian
effective actions\cite{dea}.
Our results  may, perhaps, be relevant for these areas of
research.

This paper is organized as follows: (a) We derive a new form of the
Wilson renormalization group equation; (b) We find a change of
co\"ordinates that `improves' the form of this equation, making it
closer to gradient flow; and (c) We discuss our results, and compare
to previous results on the character of renormalization group flow.
\def\D{{\hbox{D}}}
\def\cF{{\cal F}}
\def\cA{{\cal A}}
\def\Lam{\Lambda}
\def\dL#1{\Lam{{{\rm d}{#1}}\over {{\rm d}\Lam}}}
\def\pL#1{\Lam{{\partial{#1}}\over {\partial\Lam}}}
\def\d{{\hbox{d}}}
\def\ee{{\hbox{e}}}
\def\be{\beta}
\def\part{\partial}
\def\sect#1{\bigskip\centerline{#1}\bigskip}
\def\ssc{\scriptscriptstyle}
\def\eps{\epsilon}
\def\cw{{c_{\scriptscriptstyle W}}}
\def\cZ{{\cal Z}}
\def\cN{{\cal N}}
\def\parL{\Lambda\partial_\Lambda}
\def\tA{{\tilde A}}
\def\ta{{\tilde a}}
\def\tpart{\tilde{\partial}}
\def\tbeta{\tilde{\beta}}
\def\a{\alpha}
\def\dlangle{\langle\!\langle}
\def\drangle{\rangle\!\rangle}
\def\ie{{\it i.e.,}\ }
\def\eg{{\it e.g.,}\ }
\def\lang{\Big\langle}
\def\rang{\Big\rangle}
\def\b{{\rm\char'31}}  
\def\B{{\cal B}}
\def\g#1{{\gamma_{#1}}}

Three points must be emphasized right from the outset:
\begin{enumerate}
\item[i)]
Our analysis makes use of a particular negative definite
matrix, the Zamolodchikov `metric', \ie
the matrix of second derivatives of the free energy
density, which arises naturally in the
calculations.\footnote{In the following, we refrain from the 
temptation to
refer to this matrix as a `metric'---see (iii). Given as
the matrix of second derivatives of a function, it is clearly
not covariant under nonlinear coordinate changes.  One can {\it 
define} a
connection\cite{weinberg} by demanding that in the affine
frame\cite{df} given by our choice of co\"ordinates the metric takes
the desired form given below. However, we shall not have any need for
this sophistication in our calculation, which is local.} Our results
stated above are valid {\it only} with this particular choice
of matrix. While it seems difficult to construct another such matrix
on the space of all couplings,
with natural positivity (or negativity) properties, this apparent
difficulty may be a reflection of our lack of imagination.
\item[ii)]
Since we consider the space of all couplings, relevant and irrelevant,
as appropriate for the Wilson renormalization group,
a cutoff is present in the formula. In this context, we find it
convenient to consider the Wilson flows 
the flows
of
dimensionful couplings. Generally, however, one
is interested in flows 
for dimensionless couplings. These
two sets of  flows 
are related in a straightforward way, and
in our final discussion then, our
results will be expressed in terms of the more usual dimensionless
variables.
\item[iii)]
Our calculations are invariant only under
linear changes of the co\"ordinates we use on the space of field 
theories.
The deep and difficult questions
that arise when one attempts to interpret the `geometry'\cite{df} of 
such a
space
are beyond the scope of the present work.
Any geometric formulation of the problem must reduce to our explicit
computations.  One might hope that the explicit and physically clear
nature of our formula will lead to progress on the correct definition
of geometry on spaces of field theories.
\end{enumerate}

We start with the derivation of a new form of the Wilson
renormalization group equation.
We begin with a euclidean scalar field theory, in $d$ dimensions,
characterised by an
action which we divide as $S=S_0+S_i$. $S_0$ provides
the regulated propagator $p^{-2}K(p^2/\Lam^2)$, where $K$ is
some cut-off function, which makes a smooth transition from
1 to 0 in a narrow interval around $p^2\approx\Lam^2$.
$S_i$ describes the theory's interactions, and here, we are
free to make an advantageous choice of co\"ordinates on the space of
couplings.  A standard choice (see, {\it e.g.,}
Ref.~\cite{pol,hl})  is such that derivatives
of the free energy with respect
to these co\"ordinates give complete {({\it i.e.}, not necessarily 
connected)}
correlation functions.  Thus we write
\begin{equation}
S_i \equiv  \sum_{n=1}^{\infty} \int \prod_{i=1}^n \d p_i\
A^{(n)}(p_1,\dots,p_{n};\Lambda)\
\delta(\sum p_i)\ \prod_{j=1}^n \phi(p_j),
\label{interact}
\end{equation}
and the $A^{(n)}$ serve as the appropriate co\"ordinates.
It should be noted
that the $A^{(n)}$ are regarded as functions only on the space
$\sum p_i=0,$ and are thus functions of only $n-1$ momenta.
Derivatives of the free energy {\it density} $f$
give
\begin{equation}
{\delta\over {\delta A^{(n)}(p_j)}}f
={\lang \phi(p_1)\dots\phi(p_{n-1})\,\phi(-\sum_{i=1}^{n-1}p_i)\rang}
\label{Corre}
\end{equation}
where as
a matter of convention, the momentum conservation delta-functions have
been factored out of the correlation functions---thus, these 
correlation
functions are `correlation functions/unit volume'.
For convenience, in the following we will use the quantity
$p_n\equiv -\sum_{i=1}^{n-1}p_i$.

By the definition of the flow of couplings, we know that
\begin{equation}
0=\dL {}{\lang \phi(p_1)\dots\phi(p_n)\rang}
\label{first}
\end{equation}
when all of the arguments are momenta smaller than borderline,
\ie, $p_i^2<\Lam^2$ for $i=1,\ldots,n$.
Expanding the right-hand side, one has
\begin{equation}
0  = \pL{}
{\lang \phi(p_1)\dots\phi(p_n)\rang}
+ \sum_m\int \prod_{i=1}^{m-1}\
 dq_i\pL {}{A^{(m)}}(q_1,\dots,q_{m};\Lambda)\
 {\delta\over {\delta A^{(m)}(q_i)}}
{\lang \phi(p_1)\dots\phi(p_n)\rang}.
\label{Naive}
\end{equation}
Here we almost have gradient flow:
It is natural to define dimensionful flow functions as
$ F^{ n }(q_{i};\Lambda) \equiv \parL A^{(n)}(q_{i};\Lambda) ,$
as appear in the second term.
The next factor in this term may be written as the matrix
\begin{equation}
X_{mn}\equiv
{\delta\over {\delta A^{(m)}(q_i)}}{\delta\over {\delta 
A^{(n)}(p_j)}}f =
{\delta\over {\delta A^{(m)}(q_i)}}
{\lang \phi(p_1)\dots\phi(p_n)\rang}.
\label{X}
\end{equation}
The first term in eq.~\reef{Naive}\ may be written as
\begin{equation}
\pL {}
{\lang \phi(p_1)\dots\phi(p_n)\rang}
= {\delta\over {\delta A^{(n)}(p_j)}}\pL f.
\label{Comm}
\end{equation}
Note that eq.~\reef{Comm}\ holds because the
$A^{(n)}$ have no explicit factors of $\Lambda$---the partial
derivative on the l.h.s. of eq.~\reef{Comm}\ acts only on the
cutoff propagators.  Later in this 
paper, we
will introduce dimensionless couplings related to $A^{{(m)}}$ by
powers of $\Lambda,$ and eq.~\reef{Comm}\ will be modified
appropriately. In a slightly
more abstract notation, eq.~\reef{Naive}\ may now be written as
\begin{equation}
0=\part_n\Big(\pL f\Big)+F^mX_{mn}
\label{Naivetee}
\end{equation}
where $\part_n$ stands for the
functional derivative with respect to $A^{(n)}(p_j),$
and we introduced a summation convention for repeated indices
which includes integration over spacetime momenta.
Thus as desired, eq.~\reef{Naivetee}\ relates the flow 
functions
to a gradient on the space of couplings. One can contract
$F^n$ in eq.~\reef{Naivetee} to produce
\begin{equation}
-F^mX_{mn}F^n=F^n\part_n\left(\parL f\right).
\label{chum}
\end{equation}
Now $X_{mn}=\part_m\part_n f$
is a negative definite matrix due to
the thermodynamic concavity of the free energy, and hence naively,
eq.~\reef{chum}\  appears to ensure that the candidate $c$-function
\cite{zam}, $\parL f$, always increases\footnote{Note that our 
flow functions describe the change in the couplings with an increase
in the momentum-space cut-off, rather than a position-space cut-off.
With this convention, one expects a $c$-function to
satisfy $F^m\part_mC\ge0$ \cite{zam}.} along the renormalization
group flows associated with $F^n$.  It should be kept in mind that
$c$-functions are usually associated with beta functions, whereas as yet
we have only introduced the flow functions $F^n$. 

However, there is a problem with the preceding argument.
In eq.~\reef{Naive}, while the
$p_i$ are restricted to be less than borderline, the $q_i$ appearing
as the arguments of $A^{(m)}$ range over all values.
This implies that in eq.~\reef{Naivetee}, $X_{mn}$ is a `rectangular' 
matrix,
in that there is  an asymmetry between the two
indices of this infinite matrix.
{If} we extend eq.~\reef{Naive}\  to one which
included correlation functions with
borderline momenta, we may be able to produce
the desired result.

If one or more of the $p_i$ are borderline momenta,
eq.~\reef{Naive}\ is replaced by
\begin{eqnarray}
&&\parL {\lang \phi(p_1)\dots\phi(p_n)\rang}
+ \sum_m\int \prod_{i=1}^{m-1}
 dq_i\ \parL {A^{(m)}}(q_1,\dots,q_{m})\ {\delta\over {\delta 
A^{(m)}(q_i)}}
{\lang \phi(p_1)\dots\phi(p_n)\rang}
\nonumber\\
&&\quad= \left[\sum_{i=1}^n \pL {\ln K(p_i^2/\Lam^2)}\right]
{\lang \phi(p_1)\dots\phi(p_n)\rang}
\nonumber\\
&&\qquad-
\sum_{\hbox{pairs} (i,j)} \left[ \pL {\ln K(p_i^2/\Lam^2)}
\cdot {K(p_i^2/\Lam^2)\over p_i^2}\delta(p_i+p_j)\right]
{\lang \phi(p_1)\dots\widehat\phi(p_i)\dots
\widehat\phi(p_j)\dots\phi(p_n)\rang},
\label{NotNaive}
\end{eqnarray}
where the hats indicate the corresponding fields are removed from
the correlation functions in the last line. The new terms appearing 
on the
right of eq.~\reef{NotNaive}\ are  produced from differentiating
propagators with one or
both ends attached to external lines. These contributions arise 
because
only {\it internal} propagator changes are cancelled by the flow of 
couplings.
\def\PL#1{{\Lam\partial{#1}/\partial \Lam}}
\def\pL#1{{\Lam\partial{#1}/\partial \Lam}}

Now, we
want to rewrite these additional contributions in the form of a 
gradient.
Towards this end, note that the new terms
in eq.~\reef{NotNaive}\ may be written as
\begin{eqnarray}
&& \lang \int g(p)\phi(p){{\delta }\over\delta\phi(p)}
\prod^n_{i=1}\phi(p_i)\rang
 -{1\over 2}\lang\int {gK\over p^2}
{{\delta }\over\delta\phi(p)}{{\delta }\over\delta\phi(-p)}
\prod^n_{i=1}\phi(p_i)\rang
\nonumber\\
&&\qquad=
 \Big\langle \int g(p)\phi(p){{\delta }\over\delta\phi(p)}
\part_nS_i\Big\rangle -{1\over 2} \Big\langle \int {gK\over p^2}
{{\delta }\over\delta\phi(p)}{{\delta }\over\delta\phi(-p)}
\part_nS_i\Big\rangle,
\label{Work}
\end{eqnarray}
where $g(p)\equiv \parL{{\rm ln}K\left({p^2/\Lambda^2}\right)}.$
Now combining this expression with the notation introduced in
eq.~\reef{Naivetee}, we may write eq.~\reef{NotNaive}\ as
\begin{equation}
F^m\part_m\part_n f= -\part_n\left(\parL f\right)
+ \langle \part_n Y \rangle,
\label{NotGrad}
\end{equation}
with
\begin{equation}
Y\equiv{\cal O}S_i
=\left[\int g(p)\phi(p){{\delta }\over\delta\phi(p)}-{1\over 2}
\int {gK\over p^2}
{{\delta }\over\delta\phi(p)}{{\delta }\over\delta\phi(-p)}\right]S_i.
\end{equation}
Eq.~\reef{NotGrad}\  shows that an obstruction to gradient flow
will come from the fact that
\begin{equation}
 \langle \part_n Y \rangle \not= \part_n\langle Y \rangle.
\label{nottee}
 \end{equation}
In words, the expectation value of a gradient (with respect to the
couplings) is not the gradient of an
expectation value, or equivalently, the expectation value of an exact
form is not exact---this is  true for generic operators, not just $Y$.

Instead of eq.~\reef{nottee}, we have the useful identity
\begin{equation}
\langle {\cal O}\part_nS_i\rangle = \part_n\langle {\cal O}S_i\rangle
+\langle \part_nS_i{\cal O}S_i\rangle - \langle
\part_nS_i\rangle\langle{\cal O}S_i\rangle \equiv \part_n\langle
{\cal O}S_i\rangle
+\langle \part_nS_i{\cal O}S_i\rangle_c\,,
\label{Iden}
\end{equation}
which continues to hold when $\cal O$ is replaced by any functional
operator. Applying this identity in eq.~\reef{NotGrad}\ leads to
\begin{eqnarray}
&&F^m\part_m\part_nf -\Big\langle\part_nS_i
\int g(p)\phi(p){{\delta }\over\delta\phi(p)}
S_{i}\Big\rangle_c+{1\over 2 }\Big\langle\part_n
S_i\int {gK\over p^2}
{{\delta }\over\delta\phi(p)}{{\delta }\over\delta\phi(-p)}
S_{i}\Big\rangle_c
\nonumber\\
&&\qquad= -\part_n\left(\parL f\right)+
\part_n\Big\langle\int g(p)\phi(p){{\delta S_i}\over\delta\phi(p)}
 - {1\over 2}{gK\over p^2}
{{\delta^2S_i }\over\delta\phi(p)\delta\phi(-p)}
\Big\rangle.
\label{Mess}
\end{eqnarray}
Now, up to terms which are independent of the couplings,
\begin{equation}
\parL f=\lang\parL S_0\rang=-{1\over2}\lang\int g(p)\phi(p)
{{\delta S_0}\over\delta\phi(p)}\rang
={1\over2}\lang\int g(p)\phi(p)
{{\delta S_i}\over\delta\phi(p)}\rang
\label{uno}
\end{equation}
where a functional integration-by-parts is used in the
last step. Another useful identity is
\begin{equation}
{gK\over p^2}{{\delta S_0}\over\delta\phi(-p)}
=g(p)\phi(p).
\label{duo}
\end{equation}
Applying eqs.~\reef{uno}\  and \reef{duo}, as well
as functional integration-by-parts,
eq.~\reef{Mess}\ can be simplified to
\begin{eqnarray}
&&F^m\part_m\part_nf -\Big\langle\part_nS_i
\int g(p)\phi(p){{\delta }\over\delta\phi(p)}
S_{i}\Big\rangle_c+{1\over 2 }\Big\langle\part_n
S_i\int {gK\over p^2}
{{\delta }\over\delta\phi(p)}{{\delta }\over\delta\phi(-p)}
S_{i}\Big\rangle_c
\nonumber\\
&&\qquad=-{1\over 2} \part_n\Big\langle\int {\rm d}p{gK\over p^2}
{{\delta S_{i}}\over\delta\phi(p)}
{{\delta S_{i}}\over\delta\phi(-p)}
\Big\rangle .
\label{Result}
\end{eqnarray}
Now this equation is the first of our main results, which we rewrite
finally as
\begin{equation}
\left[F^m  +
\int A^{(m)}(p_i) \sum g(p_i) -H^{(m)}\right]\part_m\part_n f
= -{1\over 2} \part_n\Big\langle\int {\rm d}p{gK\over p^2}
{{\delta S_{i}}\over\delta\phi(p)}
{{\delta S_{i}}\over\delta\phi(-p)}
\Big\rangle ,
\label{RR}
\end{equation}
with
\begin{equation}
H^{(m)}\part_m\part_nf \equiv -{1\over 2}
\Big\langle\part_nS_i\int {\rm d}p{gK\over p^2}
{{\delta }\over\delta\phi(p)}{{\delta }\over\delta\phi(-p)}
S_{i}\Big\rangle.
\end{equation}
Explicitly,
\begin{equation}
H^{(m)}(p_i) \equiv
\sum_{\hbox{pairs}}\int {\rm d}p A^{(m+2)}(p_1,\dots,p,\dots,-p,
\dots,p_m) {gK\over p^2}.
\end{equation}
We see in eq.~\reef{RR}\ that terms linear in the couplings must be
subtracted from $F^{m}$ to obtain the desired gradient flow form,
proportional to the gradient of some $c$-function.
It is straightforward to show that these linear terms cannot be 
written in
the form of a gradient, implying that our Wilson flows are not the 
gradients of
any function.
The two linear terms subtracted from $F$ have a clear physical
significance: (i) The terms $A^{(m)}(p_i) \sum g(p_i)$ correspond
to momentum-dependent field rescaling at the cutoff scale; (ii) The 
terms
$H^{(m)}$ correspond to changes in normal ordering of higher order
interactions.

It is not difficult to show that eq.~\reef{RR}\ can be used to derive
Polchinski's equation\cite{pol}.  Further, we have checked that the
tachyon equation of motion in string theory \cite{bm}\ can be derived
from eq.~\reef{RR}, with the linear terms subtracted from $F^{m}$
playing a crucial r\^ole, thereby indicating that these linear terms
are physically important.

We turn now to our second result.
Let us reconsider the terms arising from borderline momenta in
eq.~\reef{Work}
describing the flow of the correlation functions. These
terms arise from the differentiation of the cutoff in the propagators 
attached
to external lines. The first term overcounts this contribution in the
exceptional case that a propagator has both ends connected to 
external lines,
and so the second term is included to compensate for this 
overcounting.
Since such propagators are disconnected from the remainder of the
correlation function, one might hope to eliminate such contributions
in the diagramatic expansion, and so improve the form of the flow 
equation,
by choosing new co\"ordinates on the space of couplings, \ie
reorganizing the interactions. In this effort, we write
\begin{eqnarray}
S_{i}& =&\sum_{n=1}^{\infty} \int \prod \d p_i
\ \tA^{(n)}(p_1,\dots,p_{n};\Lambda,\a)\ \delta(\sum p_i)
\nonumber\\
&&\quad\left[\ \prod_{i=1}^n \phi(p_i)
-\a\sum_{\hbox{ pairs}}
{K(p_j)\over p_j^2}\delta(p_j+p_k
)\!\prod^n_{\matrix{\ssc i=1\cr\ssc i\ne j,k\cr}}\!\phi(p_i)\right.
\label{longg}\\
&&\qquad\left.
+\a^2\sum_{\hbox{ distinct pairs}}{K(p_j)\over p_j^2}\delta(p_j+p_k)
{K(p_l)\over p_l^2}\delta(p_l+p_m)\!\!\prod^n_{\matrix{\ssc i=1\cr
\ssc i\ne j,k,l,m\cr}}
\!\!\phi(p_i)\  -\dots\ \right]
\nonumber
\end{eqnarray}
where $\a$ is an arbitrary constant, and the sums in the latter terms
only run over {\it distinct} pairs of indices.

Generalising in the obvious way the notation introduced
in eq.~\reef{Naivetee}, we may write \eg
\begin{eqnarray}
\lang\tpart_n S_{i}\rang
&=&\dlangle\phi(p_1)\dots\phi(p_n)\drangle_\a
\label{Exam}\\
&=&\lang\phi(p_1)\dots\phi(p_n)\rang
-\a\sum_{\hbox{pairs}}{K(p_j)\over p_j^2}\delta(p_j+p_k)
{\lang \phi(p_1)\dots\widehat\phi(p_j)\dots
\widehat\phi(p_k)\dots\phi(p_n)\rang}+\dots
\nonumber
\end{eqnarray}
Here the $\dlangle\dots
\drangle_\a$ indicates that certain diagrams in the usual expansion 
are
supplemented by extra $\a$ dependent
factors. In particular, diagrams with a
propagator which has both ends connected to external lines receive a 
factor
of $(1-\a)$. Hence the choice $\a=1$ will eliminate these 
contributions
altogether---the motivation for this approach.
The other place where these factors occur is in diagrams with
a propagator which has both ends connected to a single vertex. Thus
normal ordering contributions are weighted by $(1-\a)$, and would also
be eliminated with $\a=1$. The latter may be verified by writing the
relation between the new $\tA^{(n)}$ and the previous $A^{(n)}$ 
couplings
\begin{eqnarray}
&&\tA^{(n)}(p_1,\dots,p_{n};\Lambda,\a)=A^{(n)}(p_1,\dots,p_{n};\Lambda)
+\a\sum_{\hbox{pairs}}\int {K(q)\over q^2} A^{(n+2)}(p_1,\dots
q,\dots,-q,\dots,p_{n};\Lambda)
\nonumber\\
&&\qquad \ \ +\a^2\sum_{\hbox{distinct pairs}}\int
\prod_{i=1}^2 {K(q_i)\over q_i^2}
A^{(n+4)}(p_1,\dots,q_1,\dots,-q_1,\dots,q_2,\dots,-q_2,\dots,p_{n};
\Lambda)+\dots
\label{redefine}
\end{eqnarray}
So we see that with $\a=1$, the $\tA^{(n)}$ is given by $A^{(n)}$ plus
all of the normal ordering contributions of the
higher point couplings $A^{(n+2m)}$.

\def\tbeta{{\tilde F}}
The flow equation replacing eq.~\reef{NotNaive}\ for the new 
co\"ordinates is
\begin{equation}
\tpart_n\left(\parL f\right)
+ \tbeta^m\tpart_m\tpart_n f =
 \Big\langle \int g(p)\phi(p){{\delta }\over\delta\phi(p)}
\tpart_nS_i\Big\rangle -{1+\a\over 2} \Big\langle \int {gK\over p^2}
{{\delta }\over\delta\phi(p)}{{\delta }\over\delta\phi(-p)}
\tpart_nS_i\Big\rangle,
\end{equation}
where $\tbeta^m\equiv\parL\tA^{(m)}$. While one's naive
intuition may have anticipated a factor of $(1-\a)$ in the last term,
the appearance of $(1+\a)$ may be explained as follows: Representing 
the
action of $\parL$ on external legs with the functional operator
$\int g(p)\phi(p){{\delta }\over\delta\phi(p)}$ has produced precisely
the same overcounting of the contributions where
a propagator has both ends connected to external lines as before.
The problem is that the functional derivative does not act on the
explicit cutoff functions appearing in eq.~\reef{Exam}, which are 
responsible
for the $\a$ dependent cancellations. Hence
the second term must be included with the larger coefficient
of $(1+\a)$ to reproduce the $\parL$ action on the combined set
of correlation functions in eq.~\reef{Exam}\ in which these
contributions are suppressed. Of course, one could choose
$\a=-1$ to eliminate this correction term. This results because the
double counting of the functional operator is correct with this choice
of $\a$ since the diagrammatic rules for
$\dlangle\dots\drangle_{\a=-1}$ involve double counting
various contributions.

Proceeding as before the final result replacing eq.~\reef{RR}\ for 
these
co\"ordinates is
\begin{equation}
\left[\tbeta^m  +
\int \tA^{(m)}(p_i) \sum g(p_i) -(1+\a)
\tilde{H}^{(m)}\right]\tpart_m\tpart_n f
= -{1\over 2} \tpart_n\Big\langle\int {\rm d}p{gK\over p^2}
{{\delta S_{i}}\over\delta\phi(p)}
{{\delta S_{i}}\over\delta\phi(-p)}
\Big\rangle
\label{Rework}
\end{equation}
with
\begin{equation}
\tilde{H}^{(m)}(p_i) \equiv
\sum_{\hbox{pairs}}\int {\rm d}p\ \tA^{(m+2)}(p_1,\dots,p,\dots,-p,
\dots,p_m;\Lambda,\a) {gK\over p^2}.
\label{Hterm}
\end{equation}
Hence the new flow equation has essentially the same form as before
although a factor of $(1+\a)$ has appeared in front of the $\tilde{H}$
contribution.
So it seems in order to improve on the form of this equation, we must
choose $\a=-1$ which eliminates at least the second linear term on
the left-hand side. Of course, $\a=-1$ leaves the first linear term
and so we still do not have gradient flow  for 
these new couplings. However the improved form does open the option
of defining
\begin{equation}
B^m\equiv\left[\parL+\sum g(p_i)\right]\tA^{(m)}(p_i,\a=-1)
\label{newb}
\end{equation}
with which we may rewrite our result as
\begin{equation}
B^m \tilde \part_m\tilde \part_n f = -{1\over 2}
\tilde \part_n\Big\langle\int {\rm d}p{gK\over p^2}
{{\delta S_{i}}\over\delta\phi(p)}
{{\delta S_{i}}\over\delta\phi(-p)}
\Big\rangle\
\label{chummy}
\end{equation}
Thus we have produced a gradient flow for the $B^m$ functions.
Note that these $B^m$ reduce to our original flows when all of the
momenta are smaller than borderline.
While such a redefinition producing gradient flow would be possible
for any value of $\a$ (including $\a=0$, our original analysis),
the latter property only holds for $\a=-1$.

\def\b{\phi}
Now the previous results, eqs.~\reef{RR}\ and \reef{chummy}\ involve
flows for the dimensionful couplings $A^{(n)}$. Of course, one
is generally more interested in the flow of dimensionless couplings,
which are related to the $A^{(n)}$ by
\begin{equation}
A^{(n)}(p_{i};\Lambda) \equiv \Lambda^{\g n}
a^{(n)}(p_{i}/\Lambda; \Lambda)\ ,
\label{relation}
\end{equation}
where $\g n={d +n -  nd/2}$. The flows 
of these dimensionless couplings are
\begin{equation}
\b^m(p_{i}/\Lambda;\Lambda) \equiv \left(\parL +\sum_{i}p_{i}\cdot
{\part\over\part{p_{i}}}\right)a^{(m)}(p_{i}/\Lambda;\Lambda).
\label{defineb}
\end{equation}
Alternatively, one may write $\b^m(x_i;\Lam)=\parL a^{(m)}(x_i;\Lam)$.
These dimensionless flows 
are related to our original
dimensionful flows 
by
\begin{equation}
F^{m}(p_{i};\Lambda) =
\Lambda^{\g m}\left[\b^{m}\left({p_{i}\over\Lambda};\Lambda\right)
+ \bigg\{\g m
- \sum_{i}p_{i}\cdot{\part\over\part{p_{i}}}\bigg\}
a^{(m)}\left({p_{i}\over\Lambda};\Lambda\right)\right]\ .
\label{betarelation}
\end{equation}
While one could reconstruct the derivation of eq.~\reef{RR}\ working
entirely with the dimensionless couplings, it is much easier to
simply insert the above relations into the result presented there.
The revised equation is
\begin{equation}
 \left[\b^m+ \int\bigg\{\g m
+ 
\sum_{i}\left(g(p_i)-p_{i}\cdot{\part\over\part{p_{i}}}\right)\bigg\}
a^{(m)}({p_{i}}) -h^{(m)} \right]
D_m D_n f
= -{1\over 2} D_n\Big\langle\int {\rm d}p{gK\over p^2}
{{\delta S_{i}}\over\delta\phi(p)}
{{\delta S_{i}}\over\delta\phi(-p)}
\Big\rangle\ ,
\label{removall}
\end{equation}
where $D_{n}$ stands for a functional derivative with respect to
$a^{{(n)}}$, and
\begin{equation}
h^{(m)}(p_i) \equiv
\sum_{\hbox{pairs}}\int {\rm d}p\ a^{(m+2)}(p_1,\dots,p,\dots,-p,
\dots,p_m) {gK\over p^2}.
\end{equation}
So essentially the only difference between this form of the equation
and that in eq.~\reef{RR}\ is the introduction of two extra linear
terms on the left-hand side. These new contributions are nonvanishing
even when all of the momenta are smaller than borderline, a 
characteristic
which they have in common with the $h^{(m)}$, \ie the normal ordering
terms.

\def\B{\Phi}
We would also like to express eq.~\reef{chummy} in terms of
dimensionless flows.
Towards this end, note that
eq.~\reef{redefine}\ is consistent with defining dimensionless
variables as before: In the higher point terms,
each momentum integration is dimensionally $\Lambda^{d},$ the 
propagators
are $\Lambda^{{-2}},$  while $A^{(n+2m)}$ carries
$\Lambda^{\g {n+2m}}$ with $\g {n+2m}= (d+n-nd/2)+m(2-d)=\g n 
+m(2-d)$.
Hence all of the terms carry an overall factor $\Lam^{\g n}$, and we
may define new dimensionless couplings by
\begin{equation}
\tA^{(n)}(p_{i};\Lambda,\a) \equiv \Lambda^{\g n}
\ta^{(n)}(p_{i}/\Lambda; \Lambda,\a)\ .
\label{brelation}
\end{equation}
The flows 
of these couplings are
defined as in eq.~\reef{defineb}, but then in analogy with
eq.~\reef{newb}, we define
\begin{equation}
\B^m(p_{i}/\Lambda;\Lambda) \equiv \left[\parL +\sum_{i}\left(g(p_i)+
p_{i}\cdot{\part\over\part{p_{i}}}\right)\right]\ta^{(m)}
(p_{i}/\Lambda;\Lambda,\a=-1).
\label{definebb}
\end{equation}
With these definitions, eq.~\reef{chummy}\ becomes
\begin{equation}
 \left[\B^m\left({p_{i}\over\Lambda};\Lambda\right)
+ \bigg\{\g m
- \sum_{i}p_{i}\cdot{\part\over\part{p_{i}}}\bigg\}
\tilde a^{ (m)}({p_{i}\over\Lambda};\Lambda,\a=-1) \right]
\tilde D_m\tilde D_n f
= -{1\over 2} \tilde D_n\Big\langle\int {\rm d}p{gK\over p^2}
{{\delta S_{i}}\over\delta\phi(p)}
{{\delta S_{i}}\over\delta\phi(-p)}
\Big\rangle\ ,
\label{removed}
\end{equation}
where $\tilde D_{n}$ stands for a functional derivative with respect 
to
$\tilde a^{{(n)}}.$ Hence for the dimensionless variables, one still
has linear terms appearing on the left-hand side, even with $\a=-1$ 
and
the modified flows \reef{definebb}.

We have been careful throughout this paper in refraining from
referring to the Wilson flows, dimensionless or dimensionful, as beta
functions.  Beta functions are obtained from the dimensionless flows 
as
follows\cite{hl}: By considering a field rescaling, $\phi \rightarrow 
a\phi,$
we can derive a simple
identity for Green functions
\begin{equation}
n\part_{n}f + {\sum_{m}}' ma^{{(m)}}\part_{m}\part_{n}f + (1+2a^{2,2})
\part_{{2,2}}\part_{n} f =0,
\label{rescale}
\end{equation}
where $a^{{2,2}}(p)$ are the couplings corresponding to the operators 
$p^{2}\phi(p)\phi(-p)$ and the prime in the summation denotes
that $a^{2,2}(p)$ is not included in the summation.  
Eq.~\reef{rescale}
allows one to eliminate $\part_{{2,2}}$ in terms of $N\equiv\int
\phi\delta/\delta\phi$ and additional terms in the flows of all the
other couplings.  Beta functions are the sum of the flows we have 
worked with
in our calculations and these additional terms:
\begin{equation}
\beta^{m }=\B^{m} -m a^{{(m)}}\B^{2,2}(1+2a^{2,2})^{{-1}},
\label{vanish}
\end{equation}
with $\beta^{{2,2}}=\B^{{2,2}} - 2({1\over
2}+a^{2,2})\B^{2,2}(1+2a^{2,2})^{{-1}}
=0.$
It is the vanishing of these beta functions that characterizes fixed
points of the renormalization group, since at such points, the
correlation functions scale under changes of the cutoff.

Eq.~\reef{removed} appears to be at some remove
from the gradient flow conjecture\cite{wally} that has motivated much
of the work in this area of field theory, so we address now the 
question
of its consistency with known results\cite{zam,zam2,wally}.  The first
obvious point is that these results are concerned with beta functions,
whereas our results concern the Wilson flows. The flows
we have considered only reduce to   beta functions when $\B^{2,2}$
vanishes, {\it i.e.} when anomalous dimensions are negligible. 
The distinction between relevant, marginal and irrelevant operators is
defined only in the vicinity of fixed points.
In that vicinity as the cutoff is lowered, irrelevant couplings
converge exponentially (in $\ln \Lambda$)
to the fixed-point subspace, while conversely relevant couplings are
exponentially repelled.
Now first we note that the evidence for gradient flow provided by
Wallace and Zia\cite{wally}, is explicitly
predicated on masslessness, and scheme-independence---see also
\cite{dolan}.
However, the divergences produced by relevant operators, \eg
mass terms,  require
additive renormalizations in any renormalization procedure with an
explicit cutoff, and so are not subject to the usual arguments for
scheme-independence. In
perturbation theory, Zamolodchikov provided some evidence for
gradient flow in two dimensions\cite{zam2},
but restricted to models with operators which are very
close to marginal (see eq.~(3.9) in Ref.~\cite{zam2}), with the
deviation from marginality determined by the perturbation parameter. 
Thus
to our knowledge, there
is {\it no} evidence in the literature that suggests that gradient 
flow
should hold in the presence of relevant operators in a
renormalization procedure with a  physical cutoff. On the contrary,
certain calculations done in the context of specific models seem to
indicate that such operators destroy gradient flow \cite{dolanb}.

One might hope to ignore the relevant operators by focussing on
the renormalization group trajectories which are precisely tuned
to flow to a fixed point---see \cite{dolanb}. One might then examine
the variation of the remaining marginal and irrelevant couplings for 
gradient
flow. Now within perturbation theory
about the free-field point, the term $\{\dots\}$ in 
eq.~\reef{removed} is
negative for irrelevant operators and  vanishes for marginal operators
(while it is positive for relevant operators). The linear terms then
provide for the collapse of the irrelevant operators to
the fixed-point surface. Accordingly, the values of irrelevant
couplings at low energies are dictated by the values of marginal (and
relevant) couplings at low energies, to exponential accuracy.
In fact for the free-field fixed point, these couplings vanish
exponentially, and so the terms linear in the irrelevant couplings
would make an insignificant contribution in
eq.~\reef{removed}. Thus one could
interpret eq.~\reef{removed} as
implying gradient flow in an approximate sense, to all orders in
perturbation theory near the free field fixed point when relevant
couplings have been tuned to vanish.  Once again, though, we
emphasize that this is gradient flow not for the beta functions,
but for the Wilson flows we have considered.

Our equation should also be valid
in regions of the space of coupling
constants where new fixed points might be found.
Such fixed points are characterized by $\beta^{m}$ vanishing, which is
equivalent to
$\B^{m} \propto ma^{(m)}.$ 
 At such fixed
points, there is no reason to suppose that the set of scaling 
operators will be
the same as at the free field
fixed point.  In this case, the linear terms on the l.h.s. should on 
the
same footing as the gradient terms\footnote{ About the free field in
two dimensions,
we have checked the physical
relevance of the term in eq.~\reef{Hterm}
in string theory, as mentioned earlier, though in this case our
redefined couplings eliminate this term on the l.h.s.}.  It would be
interesting to ascertain their significance at the Wilson-Fisher
fixed point in the
$\epsilon$ expansion. Hence eq.~\reef{removed} does not appear to
give a gradient flow on the full space of couplings.

We emphasize that our present analysis has not taken into account
the existence of {\it redundant} operators---these would
correspond to the existence of couplings $g_r$ such that 
$\beta(g,g_r+\epsilon)
= \epsilon {\cal L}_{V_r}\beta(g,g_r),$ {\it i.e.}, to operators such
that translations in the redundant directions amount to a 
reparametrization
 of the space of couplings by a vector field $V_r.$ Clearly, this 
alone
 does not suffice to characterize redundant variables, for a
 detailed discussion, see Wegner\cite{weg}.  Na\"\i vely, ignoring
 reparametrizations, one could characterize redundant directions as
 those which leave $f$ invariant\cite{weg},
 {\it i.e.}, $\epsilon^r(g)\part_rf=0,$
 where $\epsilon^r$ is the vector of couplings corresponding to
 a particular redundant operator. In considering the
completeness of the co\"ordinates $A^{(m)},$ a point of note
is that these co\"ordinates are conjugate to the correlation
functions, and we
 stress that while $S$-matrix elements are left invariant by field
 redefinitions in functional integrals, this is not true for general
 correlation functions.

Despite finding only a limited gradient flow interpretation for
eq.~\reef{removed},
we still have a much simpler flow 
in eq.~\reef{chummy}. This equation leads to
\begin{equation}
-B^m\tilde{X}_{mn}B^n=-B^m(\tpart_m\tpart_n f)B^n=B^n\tpart_nC
\end{equation}
with
\begin{equation}
C={1\over 2} \Big\langle\int {\rm d}p{gK\over p^2}
{{\delta S_{i}}\over\delta\phi(p)}
{{\delta S_{i}}\over\delta\phi(-p)}
\Big\rangle\ .
\label{see}
\end{equation}
Now the thermodynamic concavity of the free energy ensures
that $C$ increases along the flows associated with $B^n$.
The interpretation of this increase is complicated by a number
of points: (i) The flow functions $B^n$ describe the flows
of dimensionful variables.  This means that a crucial step
of the renormalization group, rescaling the system back to its 
original size
after blocking (which {\it is} included if we work with $\beta^{n}$),
is missing. (ii) The modified flow functions \reef{newb}\ include
a field rescaling contribution for momenta at the
cutoff scale. Note, however, that these new functions reduce to our
original flow functions when the momenta are less than borderline.
(iii) The choice of $\a=-1$ produces an unusual diagrammatic
expansion. We should also note that the
invertibility of the Zamolodchikov `metric' is not clear in the
Wilsonian context. Hence given eq.~\reef{chummy},
it is not {\it a priori} obvious that one could solve
for     $B^n$ as an inverse matrix multiplying
the gradient of the $c$-function. Given all of these points, the 
significance
of eq.~\reef{chummy} remains unclear despite its appealing simplicity.

To conclude, we make some observations about this
potential $c$-function\reef{see}, which also appears in the various
forms of our flow equation\footnote{Note that these observations are 
based
on the relation to the dimensionful $B$ functions, so a comparison to
$c$-functions\cite{zam,car,lots} for dimensionless variables is only
justified for
marginal couplings, in the limit of vanishing anomalous dimensions.}.
If we assume the cutoff function $K$ is monotonic, then the
first factor in the integrand of eq.~\reef{see}\ is positive.
Hence our candidate
would be positive-definite, a desirable property for a $c$-function.
It is also interesting also to consider the relation of the gradient 
term in
eq.~\reef{Rework}\  to the free energy density.
One finds that up to terms independent
of the couplings that
\begin{equation}
C=-\parL f-{\a-1\over2}
\Big\langle\int {gK\over p^2}
{{\delta }\over\delta\phi(p)}{{\delta }\over\delta\phi(-p)}
S_{i}\Big\rangle\ .
\end{equation}
Thus for $\a=1$, \ie normal-ordered interactions,
one has simply $C=-\parL f$. This is precisely the negative of the
candidate $c$-function in the naive analysis which produced
eq.~\reef{Naivetee}. There, however, the flow was considered
for the couplings $A^{(m)},$ not our modified $\tA^{(m)}$.

\end{document}